\input{aipcheck}

\documentclass[
    ,final            
  ]
  {aipproc}

\layoutstyle{6x9}

\begin{document}

\title{Rapidity Dependence of Elliptic Flow at RHIC}


\author{E. B. Johnson for the BRAHMS Collaboration}{
  address={University of Kansas, Lawrence, KS 66045}
}

\begin{abstract}
The measured elliptic flow (v$_2$) of identified particles as a function of p$_T$ and centrality at RHIC
suggests the created medium in Au+Au collisions achieves early local thermal equilibrium that is followed 
by hydrodynamic expansion.  It is not known if the $\eta$ dependence on v$_2$ \cite{Phobos200,Star200} is 
a general feature of elliptic flow or reflects other changes in the particle spectra in going from mid-rapidity 
to foward rapidities. The BRAHMS experiment provides a unique capability compared to the other RHIC experiments 
to measure v$_2$ for identified particles over a wide rapidity range.  From Run 4 Au+Au collision at 
$\sqrt{s_{NN}}$ = 200GeV, identified elliptic flow is studied using the BRAHMS spectrometers, which 
cover 0$<$$\eta$$<$3.4.  The BRAHMS multiplicity array is used to determine the v$_2$ event 
plane for the identified particle elliptic flow and to measure the p$_T$-integrated flow for 
charged hadrons.

\end{abstract}

\maketitle

\indent{} \indent{}
The Relativisitic Heavy-Ion Collider has produced Au-Au collisions at $\sqrt{s_{NN}}$ = 200GeV in order to 
create a novel state of matter, the quark-gluon plasma (QGP), in which quarks and gluons are no longer confined \cite{QGPLetter}.  
Hydrodynamical models that assume the formation of a QGP have been used to model the behavior of the created medium 
\cite{SorgeHydro}. The data presented by the STAR, Phobos, and PHENIX collaborations in ref. \cite{Phobos200,Star200,Phenix200} 
show strong evidence that the created medium behaves as a fluid.  These studies have been limited, however, to the mid-rapidity 
region of the collision.  

\indent{}
BRAHMS studies elliptic flow as a function of the longitudinal expansion of the created medium.   The STAR and Phobos 
collaborations have shown that the p$_T$-integrated elliptic flow is monotonically dependent on pseudorapidity 
\cite{Phobos200,Star200}. Phobos reports on a ``limiting framentation'' behavior \cite{PhobosLimit}, showing a universal 
dependence of elliptic flow on $\eta$' (= $\eta$ - y$_{beam}$) as $\eta$' goes to zero for various collision energies.  
We seek to better understand this behavior by studying its rapidity and transverse-momentum dependence.

\indent{} 
Elliptic flow is directly influenced by the initial state of the system but is also 
affected by final state dynamics.  The p$_T$-integrated elliptic flow shows a centrality dependence that is attributed 
to the inital collision geometry.  The mid-rapidity elliptic flow dependence on p$_T$ has a number of signals suggesting 
an interplay between final and initial state effects.  The elliptic flow for charged-hadrons shows a strong dependence 
on p$_T$, and initial state hydrodynamical models are in good agreement with the measurement up to about 1.5 GeV/c 
\cite{Star200,Phenix200}.  At this point, the signal seems to saturate for the higher p$_T$ particles. The elliptic flow 
at intermediate p$_T$ is found to scale with the number of constituent quarks, n, (i.e. v$_2$/n vs. p$_T$/n has a universal dependence), 
which is consistent with a final state effect of quark coalescence \cite{qcoal}.  This dependence holds well for baryons and
kaons for p$_T$/n > 0.75 GeV/c, as shown in ref. \cite{Star200,Phenix200}, but the pions behave somewhat differently.
The deviation seen for pions could be due to resonance decays or the difference in mass between the pion and 
the constituent quarks.  

\indent{}
The BRAHMS experiment \cite{BRAHMSNIM} consists of two spectrometers, one covering angles near mid-rapidity (MRS) 
and one covering forward rapidities (FS).  A number of global detector arrays characterize the overall charged-particle
production.  The collision region is surrounded by arrays of silicon detectors (SMA) and scintillating tile detectors (TMA).
Three azimuthally symmetric rings of Si detectors, and one ring of tile detectors are used in the reaction-plane analysis.  
The BRAHMS azimuthally symmetric, left large-tube beam-beam counters (BBL) are also used for this analysis.

\indent{}
The basic procedure to measure the elliptic flow signal is outlined in ref. \cite{FlowMethod}. Once a reaction plane 
is determined for the five detector rings (3Si, 1 Tile, 1BBL), the $\phi$ of the particles measured in one detector 
is correlated to the reaction plane in another.  Several combinations of the detector rings were used for 
the integrated v$_2$ analysis, while p$_T$ dependent v$_2$ analysis only correlated the tile reaction plane with the 
particles identified in the spectrometers.

\indent{}
To obtain the correct v$_2$ value, signal distortions resulting from the reaction-plane resolution, the background, 
and other non-flow effects, need to be removed.  Since the BRAHMS experiment is not symmetric about $\eta$ = 0, 
the reaction-plane resolution measurement requires using three independent reaction planes.  Auto-correlations are 
removed by correlating detectors that are seperated by at least 0.2 units in $\eta$.   Normalized weights based on averages 
over many events are used to remove any anisotropic effects.  These weights are determined using minimum biased events for 
the intergated flow analysis, but all events are used in the spectrometer analysis.  To further remove any first order effects, 
the $\Sigma$w$_i$sin(2$\phi$$_i$) and  $\Sigma$w$_i$cos(2$\phi$$_i$) terms in the reaction plane calculation 
are centered to zero on average. From ref. \cite{Flatten}, the final reaction plane distribution can be completely flattened 
by taking a Fourier decomposition of the distribution over minimum bias events.  Background and geometrical effects are 
removed using GEANT simulations with a known elliptic flow signal.

\begin{figure}[b]
  \includegraphics[height=.23\textheight]{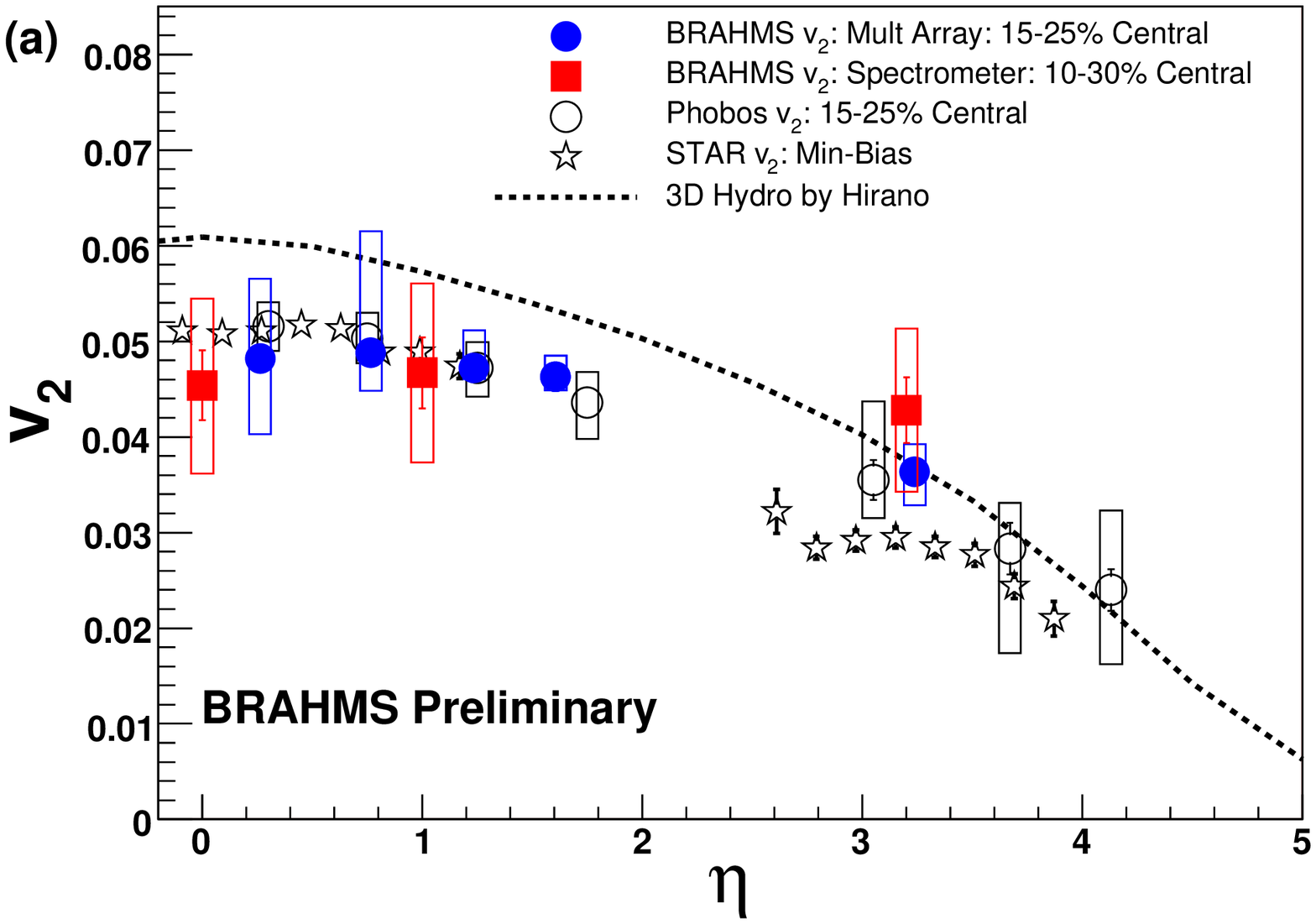}
  \includegraphics[height=.23\textheight]{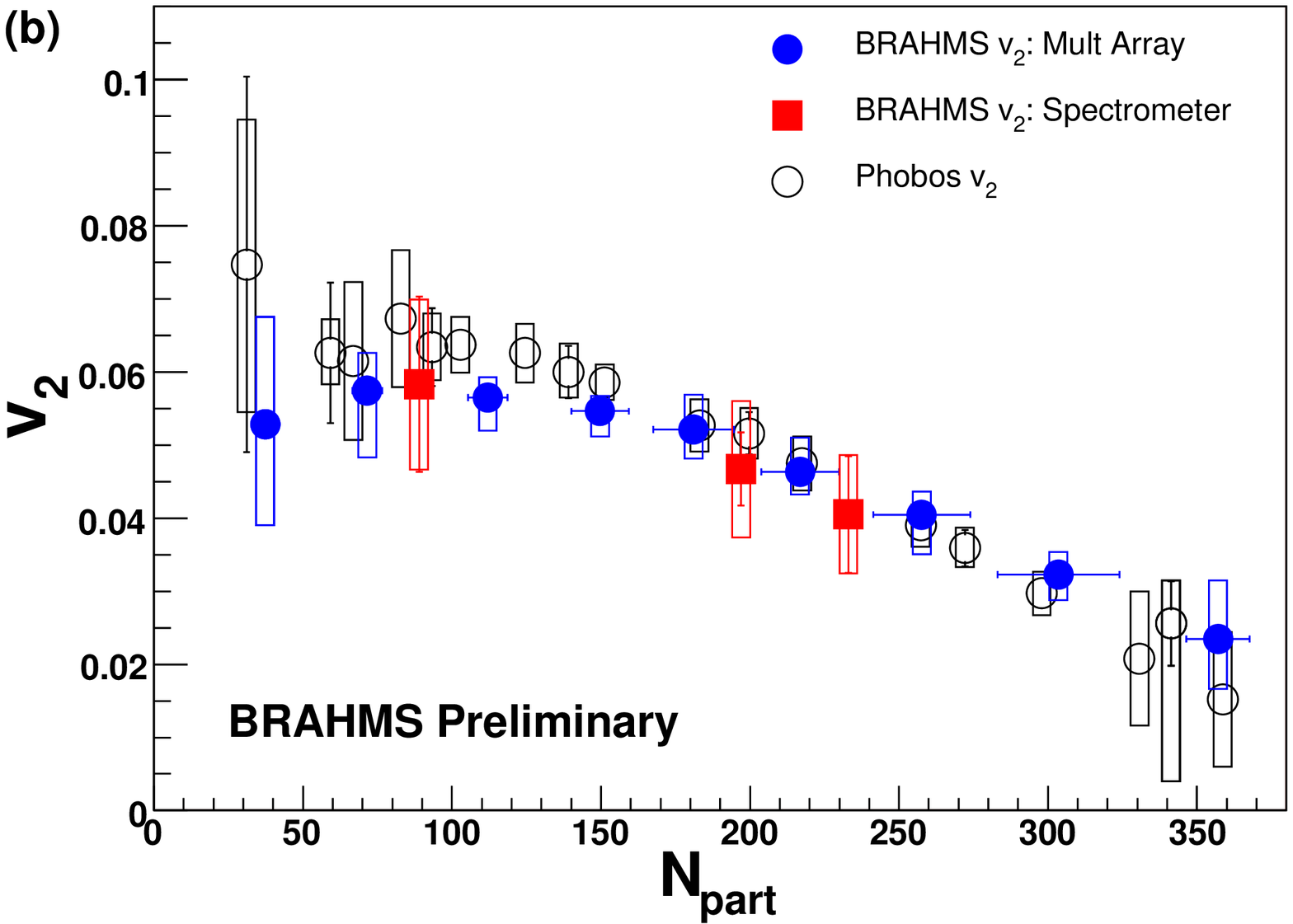}
  \caption{
	(a) Integrated v$_2$ versus $\eta$ for charged hadrons.  The BRAHMS spectrometer results are consistent, but 
	systemmatic errors with the spectra may distort the shape.  For comparison, a 3D hydro model 
	results \protect\cite{Hirano3D} are plotted. (b) Integrated v$_2$ versus N$_{Part}$ for charged hadrons within -1.0 < $\eta$ < 1.0.  
	}	
\end{figure}

\indent{}
The BRAHMS integrated v$_2$ versus centrality and $\eta$ dependencies are consistent with the Phobos results 
\cite{Phobos200} (see Fig 1a and 1b).  The p$_T$ dependent elliptic flow signal is determined for charged-hadrons 
and protons at mid-rapidity, and these results agree with the data presented by the STAR \cite{Star200} and PHENIX 
\cite{Phenix200} collaborations.  The charged-hadron spectrum at mid-rapidity is used to determine the integrated 
v$_2$ as a function of centrality for the spectrometer measurements, with results that agree within errors with the 
integrated v$_2$ determined from the BRAHMS global detectors (see Fig. 1a and 1b).  The measured charged-hadron v$_2$ 
dependence on p$_T$ for centralities from 10\% to 30\% does not seem to change much over rapidity, as shown in Fig. 2.  
This rapidity independent behavior is consistent with a 3D hydrodynamical model \cite{Hirano3D}.  The result suggests 
that the $\eta$ dependence of the integrated v$_2$ is not an inherent feature but reflects the charged-particle p$_T$ 
spectrum with rapidity.

\begin{figure}
  \includegraphics[height=.24\textheight]{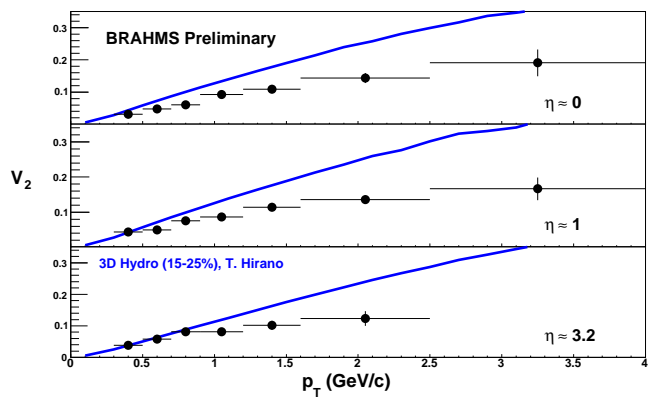}
  \caption{
	p$_T$ dependent elliptic flow for charged hadrons for the 10\% to 30\% most central events.  
	The top panel is for $\eta$ $\approx$ 0, the middle shows $\eta$ $\approx$ 1, and the bottom 
	panel displays $\eta$ $\approx$ 3.2.  The 3D hydro calculations are also given for the 15\% to 25\% 
	most central events.}
\end{figure}
 
\indent{} 
The BRAHMS experiment is able to reproduce the elliptic flow signitures seen in the other RHIC 
experiments and expand the study of identified-particle elliptic flow to forward rapidity.  The study of the 
elliptic flow over a large rapidity range will allow a better understanding of the longitudinal
expansion of the created medium. Current work is focussed on establishing the elliptic flow 
for protons, pions, and kaons over the rapidity acceptance of BRAHMS.
 
\indent{} 
Work supported in part by the Office of Nuclear Physics of US DOE under
contract DE-FG03-96ER40981 and DOE EPSCoR DE-FG02-04ER46113.

\bibliographystyle{aipprocl}

\end{document}